\documentstyle[11pt]{article}

\newcommand{\evo}{e^{\frac{i}{\hbar}tH}}
\newcommand{\evi}{e^{-\frac{i}{\hbar}tH}}
\newcommand{\snn}{\sum\limits_{n=1}^{\infty}}
\begin{document}

\title{From quantum to quantum via decoherence\thanks{\bf Dedicated to Gianfausto Dell'Antonio on his 70th birthday}}
\author{Ph. Blanchard\\
Physics Faculty and BiBoS\\ University of Bielefeld, 33615 Bielefeld \and
P. {\L}ugiewicz and R. Olkiewicz\\
Institute of Theoretical Physics\\ University of Wroc{\l}aw, 50204 Wroc{\l}aw}

\date{\today}

\maketitle

\begin{abstract}
Various physical effects resulting from decoherence are discussed in the algebraic
framework. In particular, it is shown that the environment may induce not only classical
properties like superselection rules, pointer states or even classical behavior of the
quantum system, but, what is more, it also allows the transition from statistical
description of infinite quantum systems to quantum mechanics of systems with a finite
number of degrees of freedom. It is shown that such transition holds for the quantum spin
system in the thermodynamic limit interacting with the phonon field.
\end{abstract}



\section{INTRODUCTION}
The problem of transition from microscopic to macroscopic description of Nature is a
fundamental one in the discussion of the interpretation of quantum mechanics. In recent
years decoherence has received much attention and has been accepted as the mechanism
responsible for the appearance of classicality in quantum measurements and the absence in
the real world of Schr{\"o}dinger-cat-like states \cite{Zurek2,Joos,Giul,Blan,Omnes}. It
was also shown that decoherence is a universal short time phenomenon independent of the
character of the system and reservoir \cite{Braun}. Different decoherence regimes that
are important for the experimental search of the transition between classical and quantum
worlds were discussed in \cite{Alicki}. The intuitive idea of decoherence is rather
clear: quantum interference effects for macroscopic systems are practically unobservable
because superpositions of their quantum states are effectively destroyed by the
surrounding environment. More precisely, it accepts the wave function description of such
a system but contends that it is practically impossible to distinguish between vast
majority of its pure states and the corresponding statistical mixtures. Therefore, this
approach has been called by Bell a FAPP (for all practical purposes) solution to the
measurement problem and to the Schr{\"o}dinger cat paradox. However, in spite of the
progress in the theoretical and experimental understanding of decoherence, its range of
validity and its full meaning still need to be revealed \cite{Omnes99,Raimond}.
\subsection{Algebraic framework}
Everybody agrees that concepts of classical and quantum physics are opposite in many
aspects. Therefore, in order to demonstrate how quanta become classical, it is necessary
to express them in one mathematical framework. In a recent paper \cite{blaol1} such an
algebraic framework which enables a general discussion of environmentally induced
classical properties in quantum systems has been proposed. It is worth noting that the
idea of using the same algebraic description of both quantum and classical mechanics was
suggested in \cite{Amann}. In this approach observables of any physical system are
represented by self-adjoint elements of some operator algebra $\cal M$, the so-called von
Neumann algebra, acting in a Hilbert space associated with the system. Genuine quantum
systems are represented by factors i.e. algebras with a trivial center $Z({\cal M})= {\bf
C}\cdot{\bf 1}$, $\bf 1$ stands for the identity operator, whereas classical systems are
represented by commutative algebras. Since a classical observable by definition commutes
with all other observables so it belongs to the center of algebra $\cal M$. Hence the
appearance of classical properties of a quantum system results in the emergence of an
algebra with a nontrivial center, while transition from a noncommutative to commutative
algebra corresponds to the passage from quantum to classical description of the system.
Since automorphic evolutions preserves the center of each algebra so this program may be
accomplished only if we admit the loss of quantum coherence, i.e. that quantum systems
are open and interact with their environment.

In order to study decoherence, analysis of the evolution of the reduced density matrices
obtained by tracing out the environmental variables is the most convenient strategy. More
precisely, the joint system composed of a quantum system and its environment evolves
unitarily with the Hamiltonian $H$ consisting of three parts
\begin{equation}
H\:=\:H_S\otimes{\bf 1}_E\:+\:{\bf 1}_S\otimes H_E\:+\:H_I.
\end{equation}
The time evolution of the reduced density matrix is then given by
\begin{equation}
\rho_t\;=\;{\rm Tr}_E(\evi(\rho_0\otimes\omega_E)\evo),
\end{equation}
where Tr$_E$ denotes the partial trace with respect to the environmental variables, and
$\omega_E$ is a reference state of the environment. Alternatively, one may define the
time evolution in the Heisenberg picture by
\begin{equation}
T_t(A)\;=\;P_E(\evo(A\otimes{\bf 1}_E)\evi),
\end{equation}
where $A\in {\cal M}$ is an observable of the system and $P_E$ denotes the conditional
expectation onto the algebra $\cal M$ with respect to the reference state ${\omega}_E$.
In this paper we shall work in the Heisenberg picture. Superoperators $T_t$ being defined
as the composition of a $^*$-automorphism and conditional expectation satisfy in general
a complicated integro-differential equation. However, for a large class of models, this
evolution can be approximated by a dynamical semigroup $T_t=e^{tL}$, whose generator $L$
is given by a Markovian master equation, see \cite{Ali,Omnes,Legg}. It represents on the
algebraic level irreversible evolution of the system.

We are now in a position to discuss rigorously the dynamical emergence of classical
observables. As was shown in \cite{blaol1} for each (up to some technical assumptions)
Markov semigroup $T_t$ on $\cal M$ one may associate a decomposition
\begin{equation}
{\cal M}={\cal M}_1\oplus{\cal M}_2
\end{equation}
such that both ${\cal M}_1$ and ${\cal M}_2$ are $T_t$-invariant
and the following properties hold:\\
(i) ${\cal M}_1$ is a von Neumann subalgebra of $\cal M$ and the evolution $T_t$ when
restricted to ${\cal M}_1$ is reversible, given by a one parameter group of
$^*$-automorphisms of ${\cal M}_1$.\\
(ii) ${\cal M}_2$ is a linear space (closed in the norm topology) such that for any
observable $B=B^*\in{\cal M}_2$ and any statistical state $\rho$ of the system there is
\begin{equation}
\lim\limits_{t\to\infty}<T_tB>_{\rho}\:=\:0,
\end{equation}
where $<A>_{\rho}$ stands for the expectation value of an observable $A$ in state

$\rho$.\\
The above result means that any observable $A$ of the system may be written as a sum
$A=A_1+A_2$, $A_i\in{\cal M}_i$, $i=1,\,2$, and all expectation values of the second term
$A_2$ are beyond experimental resolution after the decoherence time. Therefore, if
decoherence is efficient then almost instantaneously what we can observe are observables
contained in the subalgebra ${\cal M}_1$. In other words we apply Borel's 0th axiom:
Events with very small probability never occur. Hence all possible outcomes of the
process of decoherence can be directly expressed by the description of this subalgebra
and its reversible evolution.
\subsection{Four aspects of decoherence}
One of the effects resulting from decoherence which has been widely discussed so far is
the destruction of macroscopic interferences or, in other words, environmentally induced
superselection rules. They arise when the phase factors between states belonging to two
distinct subspaces of the Hilbert space of the quantum system are being continuously
destroyed by the interaction with its environment. The loss of quantum coherence in the
Markovian regime was established in a number of papers \cite{Unruh,Twam} giving clear
evidence of dynamical appearance of superselection rules. It was also shown that
superselection rules may emerge through the interaction of a charged particle with
electromagnetic fields \cite{GKZ}. Expressing these results in terms of the algebraic
language we will say that decoherence induces superselection rules in the quantum system
if the algebra ${\cal M}_1$ is still noncommutative but has a nontrivial center $Z({\cal
M}_1)$. Indeed, in such a case the algebra ${\cal M}_1$ is a block algebra with respect
to the decomposition of the Hilbert space ${\cal H}=\oplus{\cal H}_{\alpha}$ associated
with the central projections in ${\cal M}_1$. The discreteness or continuity of the
center $Z({\cal M}_1)$ corresponds therefore to the case of discrete or continuous
superselection rules.

Another aspect of decoherence which was analyzed in a number of models is the selection
of the preferred basis of pointer states, the so-called einselection,
\cite{Omnes,Zurek1,Dalvit}. It occurs when the reduced density matrix of the system
becomes approximately diagonal in a time much shorter than the relaxation time. Most
models predict that these states exist and are orthogonal so they allow to define a
unique set of alternative events with well definite probabilities. It follows that
pointer states do not evolve at all, while all other pure states deteriorate in time to
classical probability distributions over the one-dimensional projections corresponding to
these states. However, it should be pointed out that the algebra generated by these
projections is always of a discrete type, and, as was shown in \cite{Olk}, the
discreteness is unavoidable as long as we consider quantum systems with a finite number
of degrees of freedom. A new perspective is opened when we consider quantum systems in
the thermodynamic limit. In \cite{blaol2} it was shown that the interaction between an
infinite quantum spin system linearly coupled to a phonon field yields a selection of a
continuous family of pointer states corresponding to an apparatus with continuous
readings. These results suggest the following definition. We will say that decoherence
induces pointer states of the quantum system if ${\cal M}_1$ is commutative and the
restriction of the evolution $T_t$ to ${\cal M}_1$ is trivial, i.e. $T_t(A)=A$ for any
observable $A\in{\cal M}_1$ and all times $t$. The discreteness or continuity of the
pointer states corresponds again to the same property of the algebra ${\cal M}_1$.

The origin of deterministic laws that govern the classical domain of our everyday
experience has also attracted much attention in recent years. In particular, the
emergence of classical mechanics described by differential, and hence local, equations of
motion from the evolution of delocalized quantum states was at the center of this issue.
For example, the question in which asymptotic regime non-relativistic quantum mechanics
reduces to its ancestor, i.e. Hamiltonian mechanics, was addressed in \cite{Froh}. It was
shown there that for very many bosons with weak two-body interactions there is a class of
states for which time evolution of expectation values of certain operators in these
states is approximately described by a non-linear Hartree equation. The problem under
what circumstances such an equation reduces to the Newtonian mechanics of point particles
was also discussed in that paper. A different point of view was taken in a seminal paper
by Gell-Mann and Hartle \cite{GelH}. They gave a thorough analysis of the role of
decoherence in the derivation of phenomenological classical equations of motion. Various
forms of decoherence (weak, strong) and realistic mechanisms for the emergence of various
degrees of classicality were also presented. In the same spirit it was shown in
\cite{Lugol} that an infinite quantum system subjected to a specific interaction with
another quantum system may be effectively described as a simple classical dynamical
system. More precisely, the effective observables of the system were parameterized by a
single collective variable which underwent a continuous periodic evolution. These results
lead us to the following definition. We will say that decoherence induces classical
behavior of the quantum system if ${\cal M}_1$ is commutative and its evolution is given
by a continuous flow on the configuration space of the algebra ${\cal M}_1$.

While the interaction of quantum systems with their environment contributes a great deal
to the appearance of classical reality like superselection rules, pointer states and
classical dynamics, this is not the whole story. It is clear from the above discussion
that something is missing in the presented effects of decoherence. Indeed, it may happen
that phase factors are destroyed in such a specific way that the observables immune to
decoherence form again a noncommutative algebra with a trivial center. In such a case,
which, as far as we know, has never been addressed, one may speak of the appearance of a
new genuine quantum system without any classical properties and with completely different
quantum properties. The most interesting example of such an effect is of course the
reduction of an infinite quantum system to a quantum system possessing only one degree of
freedom. This would help in the understanding how it is possible that quantum mechanics
is so efficient in the world, where almost all quantum objects should be described in
terms of quantum field theory. The possibility of such transition is the main objective
of the present paper. For its derivation we consider a completely solvable but simplified
model of an infinite array of spin-$\frac{1}{2}$ particles. Since we neglect the position
variables what we achieve is a toy model of quantum mechanics represented by a spin
algebra of $2\times 2$ matrices with the Hamiltonian evolution given by the third Pauli
matrix. This simplified model suggests, however, the possibility of deriving the
Schr{\"o}dinger equation form quantum theory of infinite systems interacting with their
environment.
\section{DECOHERENCE INDUCED SPIN ALGEBRA}
There are two approaches to the algebraic structure associated with a quantum system. In
the first one one starts with the Hilbert space of states of the system and subsequently
introduces the algebra of operators corresponding to physical observables. In the second
approach of statistical mechanics one postulates certain structural features, like
canonical commutation or anticommutation relations, of an abstract algebra, and then
recovers the traditional point of view by passing to a particular representation, the
so-called Gelfand-Naimark-Segal (GNS in short) representation, of the algebra
\cite{brat}. Clearly, the description of quantum systems in the thermodynamic limit by
statistical mechanics is an idealization of a finite physical system with a huge number
of degrees of freedom by an infinite theoretical model. Nevertheless, such an approach
proved to be very efficient in many concrete problems. In this section we use this
algebraic framework to discuss the transition of an infinite system of spin-$\frac{1}{2}$
particles, linearly coupled to a phonon field, to the spin algebra.
\subsection{The model}
The infinite quantum spin system consists of a set of noninteracting spin-$\frac{1}{2}$
particles fixed at positions $n=1,2,...$ and exposed to a magnetic field. The algebra
$\cal M$ of its bounded observables is given by the $\sigma$-weak closure of
$\pi_0(\otimes_1^{\infty}M_{2\times 2})$, where $\pi_0$ is a (faithful) GNS
representation with respect to a tracial state tr on the Glimm algebra
$\otimes_1^{\infty}M_{2\times 2}$, and $M_{2\times 2}$ is the algebra generated by Pauli
matrices. Let us point out that $\cal M$ is not a "big" matrix algebra. It is a
continuous algebra (factor of type II$_1$) in which there are no pure states. In fact,
any projection $e\in{\cal M}$ contains a nontrivial subprojection $f\in{\cal M}$. It is
worth noting that the absence of minimal projections is a new feature which may be
present only in systems in the thermodynamic limit. Since the particles are
noninteracting, their evolution is given by a free Hamiltonian which corresponds to the
interaction of the spins with an external magnetic field parallel to the $z$-axis and of
strength $H(n)$ at the site $n$
\begin{equation}
H_S\:=\:\pi_0\left(-g\mu_B\snn H(n)\sigma_n^3\right),
\end{equation}
where $g$ is the Land\'{e} factor, $\mu_B$ is the Bohr magneton and $\sigma_n^3$ is the
third Pauli matrix in the $n$th site. We assume that the magnetic field decreases as
$H(n)\sim(\frac{1}{q})^n$ for some $q\geq 2$. Since the coefficients $H(n)$ are summable,
the Hamiltonian $H_0$ is bounded. Moreover, its eigenvalues are nondegenerate.

The reservoir is chosen to consist of noninteracting phonons of an infinitely extended
one dimensional harmonic crystal at the inverse temperature $\beta =\frac{1}{kT}$. The
Hilbert space $\cal H$ representing pure states of a single phonon is (in the momentum
representation) ${\cal H}=L^2({\bf R},\,dk)$. A phonon energy operator is given by the
dispersion relation $\omega(k)=|k|$ ($\hbar =1,\;c=1)$. It follows that the Hilbert space
of the reservoir is ${\cal F}\otimes{\cal F}$, where $\cal F$ is the symmetric Fock space
over $\cal H$. A phonon field $\phi(f)=\frac{1}{\sqrt{2}}(a^*(f)+a(f))$, where $a^*(f)$
and $a(f)$ are given by the Araki-Woods representation \cite{Araki}:
\begin{equation}
a^*(f)\:=\:a_F^*((1+\rho)^{1/2}f)\otimes I\:+\:I\otimes a_F(\rho^{1/2}\bar{f}),
\end{equation}
\begin{equation}
a(f)\:=\:a_F((1+\rho)^{1/2}f)\otimes I\:+\:I\otimes a_F^*(\rho^{1/2}\bar{f}).
\end{equation}
Here $a_F^*(a_F)$ denotes respectively creation (annihilation) operators in the Fock
space, and $\rho$ is the thermal equilibrium distribution related to the phonons energy
according to the Planck law
\begin{equation}
\rho(k)\:=\:\frac{1}{e^{\beta\omega(k)}-1}.
\end{equation}
Since the phonons are noninteracting, their dynamics is completely determined by the
energy operator
\begin{equation}
H_E\:=\:H_0\otimes I\:-\:I\otimes H_0,
\end{equation}
where $H_0=d\Gamma(\omega)=\int\omega(k)a_F^*(k)a_F(k)dk$ describes dynamics of the
reservoir at zero temperature. The reference state of the reservoir is taken to be a
gauge-invariant quasi-free thermal state given by
\begin{equation}
\omega_E(a^*(f)a(g))\:=\:\int\rho(k)\bar{g}(k)f(k)dk.
\end{equation}
Clearly, $\omega_E$ is invariant with respect to the free dynamics of the environment.

The Hamiltonian $H$ of the joint system consists of the three parts $H=H_S+H_E+H_I$,
where $H_I$ is the interacting Hamiltonian. We assume that the coupling is linear (as in
the spin-boson model), i.e. $H_I=\lambda Q\otimes\phi(g)$, where
\begin{equation}
Q\:=\:\pi_0\left(\sum\limits_{n=1}^{\infty}a_n\sigma_n^1\right),
\end{equation}
$\sigma_n^1$ stands for the first Pauli matrix in the $n$th site, $\lambda>0$ is a
coupling constant, and $a_n\sim(\frac{1}{p})^n$ for some $p\geq 2$. Again, since the
coefficients $a_n$ are summable, the coupling operator $Q$ is bounded and has a
nondegenerate spectrum. Finally, we impose some restriction on the test function $g(k)$
of the phonon field. We assume that $g(k)=|k|^{1/2}\chi(k)$, where $\chi(k)$ is an even
and real valued  function such that: (i) $\chi$ is differentiable with bounded
derivative, (ii) for large $|k|$, $|\chi(k)|\leq\frac{C}{k^{2+\epsilon}}$, $C>0$,
$\epsilon>0$, and $\chi(0)=1$. The behavior of the test function $g$ at the origin and
its asymptotic bound are taken to ensure that $H$ is essentially self-adjoint. Hence it
induces a unitary evolution of the compound system.
\subsection{Description of effective observables}
The reduced (irreversible) dynamics of the system is given by Eq. (3) with the
Hamiltonian $H$ introduced in the previous subsection. Because neither $H_S$ and $H_I$,
nor $H_E$ and $H_I$ commute, it is a nontrivial step to derive an explicit formula for
the superoperators $T_t$. However, as was mentioned in the Introduction, one may apply
the Markovian approximation to simplify the problem. Because the thermal correlation
function
\begin{eqnarray}
<\phi_t(g)\phi(g)>\:=\:\omega_E(e^{itH_E}\phi(g)e^{-itH_E}\phi(g))\nonumber\\
=\:\omega_E(\phi(e^{it\omega}g)\phi(g))
\end{eqnarray}
is integrable, we use the so-called singular coupling limit to conclude that $T_t=e^{tL}$
is a quantum Markov semigroup with the generator $L$ given by the following master
equation, see \cite{blaol2},
\begin{equation}
L(A)\:=\:i[H_S-bQ^2,\,A]\:+\:L_D(A),
\end{equation}
where
\begin{equation}
L_D(A)\:=\:\frac{2\pi\lambda}{\beta}(QAQ\:-\:\frac{1}{2}\{Q^2,\,A\}),
\end{equation}
and $b=\int_0^{\infty}\chi^2(k)dk>0$. The first part in Eq. (14) is the commutator with a
new collective Hamiltonian $H_C=H_S-bQ^2$, while the second term is a dissipative
operator. The collective Hamiltonian
\begin{eqnarray}
H_C\:=\:-\pi_0\left(g\mu_B\snn H(n)\sigma_n^3\right)\nonumber\\
-\:\pi_0\left( b\sum\limits_{n,m=1}^{\infty}J(n+m)\sigma_n^1\sigma_m^1\right),
\end{eqnarray}
where $J(n)=a_n$, is similar to that of the anisotropic Heisenberg model with an infinite
range interaction. However, the potentials $H(n)$ and $J(n)$ are not translationally
invariant.

We are now in a position to formulate our main result (its proof will be given
in the Appendix).\\
THEOREM: {\it For the semigroup $T_t=e^{tL}$ the decomposition (4) holds with ${\cal
M}_1={\bf C}\cdot{\bf 1}_S$. If we put in Eq. (12) $a_1=0$, then ${\cal M}_1=M_{2\times
2}$ and for any} $A\in{\cal M}_1$
\begin{equation}
T_t(A)\:=\:e^{ith_1\sigma^3}Ae^{-ith_1\sigma^3},
\end{equation}
{\it where} $h_1=H(1)$.\\
This result shows that the infinite quantum spin system, subjected to a specific
interaction with the phonon field, after the decoherence time may be effectively
described as a quantum system with only one degree of freedom (generalization to a finite
number of degrees of freedom is straightforward). In other words, the environment forces
the spin particles to behave in a collective way what allows introduction of three
collective observables which satisfy the standard commutation relations of spin momenta.
Although, the presented model neglects position variables and so is not complex enough to
allow derivation of the Schr{\"o}dinger equation, it suggests that decoherence induced
reduction of quantum statistical mechanics of many body systems to quantum mechanics of
wave functions is possible.

\appendix*
\section{Proof of theorem}
Step 1. It is clear from the form of the generator $L$, see Eq. (14), that it generates a
semigroup of completely positive and normal superoperators which are contractive in the
operator norm. Moreover, tr$L(A)=0$, which implies that the semigroup $T_t$ is trace
preserving. Hence the decomposition (4) follows from Theorem 11 in \cite{blaol1}.

Step 2. The subalgebra ${\cal M}_1$ is defined by the property $T_t^*T_tx=T_tT_t^*x=x$
for all $t\geq 0$ \cite{Lugol}. Hence
$$
\bigcap_{l=0}^{\infty} \hbox{ker}L_D \circ \delta_{H_C}^l \subset {\cal M}_{1},
$$
where $\delta_{H_C}(\cdot)=i[H_C,\,\cdot]$. We prove now the reverse inclusion.
Suppose that $x\in{\cal M}_1$. Then, by differentiating the equation $T_t^*T_tx=x$ at
time $t=0$, we get ${\cal M}_1\subset\mbox{ker}L_D$. Assume that
$$
{\cal M}_{1} \subset \bigcap_{l=0}^{n-1}\hbox{ker}L_D \circ \delta_{H_C}^l
$$ 
for some $n\geq 1$. Because
$$\frac{d^{n+1}}{dt^{n+1}}T_t^*T_tx|_{t=0}\:=\:0$$ 
so
\begin{eqnarray}
\lefteqn{\frac{d^{n+1}}{dt^{n+1}}T_t^*T_tx|_{0}=}\nonumber\\
&=& (-\delta_{H_C} +L_D)^{n+1}(x) + \sum_{m=1}^n
\left({n+1 \atop m}\right)(-\delta_{H_C} + L_D)^{n+1-m}
\circ (\delta_{H_C}+L_D)^m(x)\nonumber\\
& & + (\delta_{H_C} +L_D)^{n+1}(x)\nonumber\\
&=& (-1)^{n+1}\delta_{H_C}^{n+1}(x) + (-1)^nL_D\circ
\delta_{H_C}^{n}(x) + \delta_{H_C}^{n+1}(x) + L_D\circ \delta_{H_C}^{n}(x)
+ \nonumber\\
& & \sum_{m=1}^n \left({n+1 \atop m}\right)(-\delta_{H_C} + L_D)^{n+1-m} \circ
\delta_{H_C}^m(x)\nonumber\\
&=& (-1)^{n+1}\delta_{H_C}^{n+1}(x) + (-1)^nL_D\circ \delta_{H_C}^{n}(x) +
\delta_{H_C}^{n+1}(x) + L_D\circ \delta_{H_C}^{n}(x) + \nonumber\\
& & \sum_{m=1}^n \left({n+1 \atop
m}\right)(-1)^{n+1-m}\delta_{H_C}^{n+1}(x)\nonumber\\
& & + \sum_{m=1}^n \left({n+1 \atop m}\right)(-1)^{n-m} L_D \circ \delta_{H_C}^n(x)
= \delta_{H_C}^{n+1}(x)\sum_{m=0}^{n+1} \left({n+1 \atop m}\right)(-1)^{n+1-m}\nonumber\\
& & + \{1+ (-1)^n[ \sum_{m=0}^{n+1} \left({n+1 \atop m}\right)(-1)^{m}-(-1)^{n+1}] \} L_D
\circ \delta_{H_C}^{n}(x) = 2L_D \circ \delta_{H_S}^{n}(x) = 0.\nonumber
\end{eqnarray}
Hence, by induction,
$$
{\cal M}_{1} \subset \bigcap_{l=0}^{\infty}\hbox{ker}L_D \circ \delta_{H_C}^l.
$$

Step 3. Let $C_1$ (respectively $C_3$) be a $C^*$-subalgebra in the Glimm algebra
generated by $\{\sigma_1^{i_1}...\sigma_n^{i_n}\}$, where $i_k=0,\,1$ ($i_k=0,\,3$
respectively), and $n\in{\bf N}$. Then both $\pi_0(C_1)$ and $\pi_0(C_3)$ are maximal
Abelian self-adjoint algebras (m.a.s.a in short) in $\cal M$ such that $\pi_0(C_1)\cap
\pi_0(C_3)={\bf C}\cdot{\bf 1}_S$. The choice of coefficients $(H(n))$ and $(a_n)$
guarantees that $L^{\infty}(Q)=\pi_0(C_1)$ and $L^{\infty}(H_S)=\pi_0(C_3)$, where
$L^{\infty}(Q)$ is the von Neumann algebra generated by operator $Q$. Hence
$L^{\infty}(Q)\cap L^{\infty}(H_S)={\bf C}\cdot {\bf 1}_S$.

Step 4. We show now that if $[Q,\,[Q,\,x]]=0$ for some $x\in{\cal M}$, then $x\in
L^{\infty}(Q)$. Let us define the derivation $\delta_x(\cdot)=i[\cdot ,\,x]$. If
$[Q,\,[Q,\,x]]=0$, then $[Q,\,x]\in L^{\infty}(Q)$ since, by step 3, $L^{\infty}(Q)$ is a
m.a.s.a. Suppose that $P$ is a polynomial. Then
$$
\delta_x(P(Q))\:=\:i[Q,\,x]P'(Q)\in L^{\infty}(Q).
$$
This implies that $\delta_x(L^{\infty}(Q))\subset L^{\infty}(Q)$ since $\delta_x$ is
continuous in the weak operator topology. Because $L^{\infty}(Q)$ is commutative so
$\delta_x|_{L^{\infty}(Q)}=\:0$, and hence $[Q,\,x]=0$. Because $L^{\infty}(Q)$ is a
m.a.s.a so $x\in L^{\infty}(Q)$.

Step 5. Next we show that ker$L_D\cap L^{\infty}(H_C)'={\bf C} \cdot {\bf 1}_S$. Here
$L^{\infty}(H_C)'$ stands for the commutant in $\cal M$ of the algebra $L^{\infty}(H_C)$.
Suppose that $x\in\mbox{ker}L_D\cap L^{\infty}(H_C)'$. Then $[Q,\,[Q,\,x]]=0$ and
$[H_C,\,x]=0$. By step 4, $x\in L^{\infty}(Q)$ which implies that $[H_S,\,x]=
[H_C+bQ^2,\,x]=0$. Hence $x\in L^{\infty}(H_S)$. Because, by step 3, $L^{\infty}(Q)\cap
L^{\infty}(H_S)={\bf C}\cdot {\bf 1}_S$ so $x=z{\bf 1}_S$, where $z\in{\bf C}$.

Step 6. By step 2, $\delta_{H_C}({\cal M}_1)\subset{\cal M}_1$.  Hence, the derivation
$\delta_1:=\delta_{H_C}|_{{\cal M}_1}$ is well defined and bounded. Thus $\delta_1(\cdot
)=i[H_1,\cdot ]$, where $H_1=H_1^*\in{\cal M}_1$ \cite{Sakai}. By step 2 again,
$H_1\in\mbox{ker}L_D$. On the other hand
$$
[H_C,\,H_1]\:=\:-i\delta_1(H_1)\:=\:[H_1,\,H_1]\:=\:0.
$$ 
Hence $H_1\in
L^{\infty}(H_C)'$ and so, by step 5, $H_1$ is proportional to the identity operator.
Suppose now that $x\in{\cal M}_1$. Then $[H_C,\,x]=-i\delta_1(x)=0$, and so 
$x\in L^{\infty}(H_C)'$. Because $x\in\mbox{ker}L_D$ so, by step 5, $x$ is proportional to the
identity operator. Hence ${\cal M}_1={\bf C}\cdot{\bf 1}_S$.

Step 7. Finally, suppose that in Eq. (12) the coefficient $a_1=0$. The corresponding
semigroup we shall denote by $T_t^1$. Let $\cal A$ be a subalgebra in $\cal M$ generated
by $\{\pi_0(\sigma_1^k):\; k=0,1,2,3\}$. Suppose that $x\in{\cal M}$. Then
$$
x\:=\:\sum\limits_{k=0}^3\pi_0(\sigma_1^k)x_k,
$$
where operators $x_k$ belong to ${\cal
A}'$, the commutant in $\cal M$ of algebra $\cal A$. Let $S_t$ be a semigroup on $\cal M$
with a generator $L_0$ given by the following Markov master equation
$$
L_0(A)\:=\:i[H_S^0-bQ^2,\,A]\:+\:L_D(A),
$$ 
where $L_D$ is defined in Eq.
(15), and 
$$
H_S^0\:=\:\pi_0\left(-g\mu_B\sum\limits_{n=2}^{\infty}
H(n)\sigma_n^3\right).
$$ 
Note that the summation index ranges from 2 to infinity. Then
$$
T_t^1(x)\:=\:\sum\limits_{k=0}^3\pi_0(U_t^*\sigma_1^kU_t)S_t(x_k),
$$ 
where
$U_t=e^{-ith_1\sigma_1^3}$. Let $L^2({\cal M})$ be the noncommutative Hilbert space of
square integrable (with respect to the trace tr) operators. Since operators $\pi_0(U_t^*
\sigma_1^kU_t)$, $k=0,1,2,3$, are orthogonal in $L^2({\cal M})$ so
$$
\|T_t^1(x)\|_{L^2}^2\:=\:\sum\limits_{k=0}^3\|S_t(x_k)\|_{L^2}^2.
$$ 
Let us notice that
the semigroup $S_t$ restricted to the commutant ${\cal A}'$ has the same properties as
the semigroup $T_t$. Hence, if any of $x_k$ is not proportional to the identity operator,
then, by step 6, $\|S_t(x_k)\|_{L^2}<\|x_k\|_{L^2}$ for all $t>0$. Thus
$\|T_t^1(x)\|_{L^2}<\|x\|_{L^2}$, too, which implies that such an operator cannot belong
to ${\cal M}_1$. Hence, if $x\in{\cal M}_1$, then $x_k=z_k{\bf 1}_S$, $z_k\in{\bf C}$,
for all $k=0,1,2,3$, and so $x\in{\cal A}$. It follows that ${\cal M}_1={\cal
A}=M_{2\times 2}$, and the dynamics on it is given by unitary operators $U_t$.

\end{document}